# Reconfigurable *p-n* Junction Diodes and the Photovoltaic Effect in Exfoliated MoS$_2$ Films

Surajit Sutar[1], Pratik Agnihotri[1], Everett Comfort[1], T. Taniguchi[2], K. Watanabe[2], and Ji Ung Lee[1*]

[1]The College of Nanoscale Science and Engineering (CNSE), SUNY at Albany, Albany, NY-12203, USA

[2]National Institute of Materials Science, 1-2-1 Sengen, Tsukuba-city Ibaraki 305-0047, Japan

**Realizing basic semiconductor devices such as *p-n* junctions are necessary for developing thin-film and optoelectronic technologies in emerging planar materials such as MoS$_2$. In this work, electrostatic doping by buried gates is used to study the electronic and optoelectronic properties of *p-n* junctions in exfoliated MoS$_2$ flakes. Creating a controllable doping gradient across the device leads to the observation of the photovoltaic effect in monolayer and bilayer MoS$_2$ flakes. For thicker flakes, strong ambipolar conduction enables realization of fully reconfigurable *p-n* junction diodes with rectifying current-voltage characteristics, and diode ideality factors as low as 1.6. The spectral response of the photovoltaic effect shows signatures of the predicted band gap transitions. For the first excitonic transition, a shift of $>4_{k_BT}$ is observed between monolayer and bulk devices, indicating a thickness-dependence of the excitonic coulomb interaction.**

Two-dimensional (2-D) crystalline materials have attracted a significant amount of research efforts since the isolation of graphene by micromechanical exfoliation [1, 2, 3, 4]. They show promise in novel electronic and optoelectronic applications, where the low-dimensionality provides ideal electrostatic control for field-effect transistor devices, or large area-to-volume ratio for sensors and photoelectric devices. Among 2-D crystals, MoS$_2$, a transition metal dichalcogenide (TMDC), has received particular attention as channel material for thin-film or flexible electronics [5, 6, 7] because its mobility is considerably higher than amorphous or polycrystalline materials, and because it can be used in various heterostructures to enable diverse electronic applications [8, 9, 10, 11, 12].

The most remarkable attributes of MoS$_2$ lie in its bandstructure, which shows a crossover from an indirect bandgap (~1.3 eV) in bulk to a direct one (~1.9 eV) for a monolayer [13, 14]. In the monolayer form, MoS$_2$ has been used in optoelectronics [15, 16, 17], and has been investigated to enable a new class of devices for emerging technologies such as valleytronics [18], [19], [20], [21]. In addition, related TMDCs with very similar lattice constants but different bandstructures [17] open the possibility of building heterostructures for efficient detection, harvesting, or generation of light over a wide spectrum, ranging from infrared to visible light. Since many optoelectronic devices require a built-in electric field for their proper function, we have set-out to fabricate and characterize *p-n* junction diodes in exfoliated $MoS_2$ films. In addition, as the basic building block of all modern semiconductor electronics, the study of *p-n* junction in any new semiconductor material can reveal previously unexplored materials properties. The formation of *p-n* junctions in a few-layer $MoS_2$ has been reported by others, using chemical doping by plasma treatment [22] or asymmetric bias between contacts relative to an ionic gate [23, 24]. In this paper, we present a robust technique to form controllable, reconfigurable *p-n* junctions in $MoS_2$ films using a pair of buried split-gates (SG).

---

[*] Corresponding author: jlee1@albany.edu

The SGs are fabricated from 100 nm thick patterned polysilicon buried under 100 nm thick $SiO_2$ in a process described elsewhere [25]. The SGs are arranged in the form of interdigitated fingers over a large area to allow mechanical exfoliation and detection of 2-D crystals, with the spacing between the SG ranging from 100 to 200 nm. Single crystals of $MoS_2$ (SPI supplies) are mechanically exfoliated on the top surface; thin layers are identified optically and characterized by AFM and Raman measurements to determine their layer thickness [26]. Electrical contacts to the $MoS_2$ flakes are defined by electron beam lithography, followed by electron beam evaporation of contact metal and lift-off processes. We examined different work-function metals to find the optimal contacts to $MoS_2$ films. To reduce the influence from defects and charged impurities at the $SiO_2$ interface, we exfoliated single crystal h-BN before exfoliating $MoS_2$ on top. Individual $MoS_2$ flakes on h-BN were identified optically and confirmed by Raman measurements [26].

To characterize the $MoS_2$ devices, we first examine the field-effect transfer characteristics of the devices by sweeping the biases on the buried gates $V_{G1}$ and $V_{G2}$ together while keeping a fixed bias $V_{DS} =$ 100 mV between the source (S) and drain (D) electrical contacts, as shown in Fig. 1; the inset shows a schematic of the device structure. A distinguishing feature in the transfer characteristics, regardless of flake thicknesses, is the stronger *n*-type conduction compared to the *p*-type conduction, also reported by others [5, 27]. The *p*-type conduction weakens with reduced flake thickness and disappears entirely for bilayer and monolayer devices. We note that the asymmetry seen in Fig. 1 could arise from gating of the contacts, since we are unable to distinguish this from channel modulation in a two-terminal measurement. Asymmetry in the *p*- and *n*-conduction has also been observed in other low-dimensional materials, including semiconducting carbon nanotubes (CNTs) [28, 29]. This asymmetry has been attributed to the modulation of Schottky tunnel barrier [30, 31, 32] or unintentional doping by adsorbates [33, 30]. Our results indicate that both effects are present in our devices since (a) we observe a thickness dependent transition from ambipolar to unipolar conduction, which we attribute to the thickness dependent bandstructure of $MoS_2$ affecting the Schottky contact barrier height, and (b) defect trap energy states due to adsorbates modify our transport properties, as we discuss below.

To better understand the cause of the asymmetry in our $MoS_2$ devices, we investigated different work-function metals. We examined Ti, Mo, Cr, Ni, and Pd with work functions that range from 4.2 to 5.4 eV. In addition, we chose to examine h-BN as an alternate substrate to $SiO_2$ because it is known to produce high mobility devices on graphene [34]. Despite the wide range in values for the work function of these metals, for flakes only a few layers thick we observed little change in the *p*-conduction, which remained negligible. For flakes of moderate thickness (15-20 nm), on the other hand, we observed *p*-conduction for all the metal contacts, which were comparable to each other and strongest for Mo (20 nm) capped by Au (30 nm), the contacts used for all the devices presented in this paper. We didn't observe significant variation in the *n*-conduction. In a previous report [35], the authors suggest that the chemical interaction between the metal and $MoS_2$, rather than the metal work function itself, had a stronger influence on conduction in $MoS_2$ (for Au and Pd, two metals with similar work functions, only *n*-, or *p*-type conduction, respectively, were observed in 50 nm thick $MoS_2$ flakes). As a comparison, flakes with a similar thickness in our work, e.g. the 62 nm thick device in Fig. 1 (Mo contacts), show fully ambipolar conduction. These differences may arise from, as mentioned in [35], actual shifts in the Fermi level at the metal-MoS$_2$ interface, which in turn are influenced by trap states due to substrate defects and adsorbates. This explanation is also consistent with an improvement in the *p*-conduction we observe in our study for moderately thick flakes fabricated on h-BN flakes, e.g. the 18 nm thick device 0668 in Fig. 1, compared to those placed directly on SiO$_2$. This device is the only one among all presented in this paper to be placed on h-BN (~50 nm thick), and shows a *p* conduction strong enough to allow creating *p-n* junctions with rectifying *I-V* characteristics, as will be discussed later. On the other hand, though similarly thick MoS$_2$ devices placed on SiO$_2$ showed *p* conduction, it was not strong enough to lead to significant rectification from *p-n* junctions created in these devices. We therefore surmise that substrate-induced impurities and adsorbates had played a dominant role, perhaps by pinning the Fermi level at interface states. In Fig. 1, we also observe that the minimum current for thicker flakes is considerably higher than that of the monolayer or bilayer device, which may imply that at distances far from the MoS$_2$-dielectric interface, gate control of the electrostatic potential and carrier modulation is not effective, and a residual carrier density contributes to the drain-to-source current ($I_{DS}$) irrespective of the gate voltage. This results in a shunt resistance that becomes more pronounced under illumination, as we discuss later in this text.

The ambipolar conduction in thicker flakes makes possible the formation of *p-n* junction diodes with appropriate biases to the SG, as shown in Fig. 2(a). Here, the current-voltage ($I_{DS}$-$V_{DS}$) characteristics from two devices are shown (marker traces). For each device, the SG biases lie on either side of the voltage at the minimum in the transfer curve, to create a *p-n* doping profile across the channel. The creation of *p-n* junctions is evident in the main plot with the observation of rectifying $I_{DS}$-$V_{DS}$ characteristics, the characteristic feature for any diode. The forward bias characteristics follow an exponential dependence with voltage for several decades, before being limited by a series resistance, which includes, apart from the contact resistances, the resistances of the *p* and *n* regions away from the *p-n* junction. To demonstrate that our diodes are reconfigurable, we switch the biases used on the SG in Fig.

2(a), to change the diode to an *n-p* doping configuration. The resulting $I_{DS}$-$V_{DS}$ characteristics are plotted in Fig. 2(b), which are mirror images to the characteristics seen in the *p-n* configuration.

The $I_{DS}$-$V_{DS}$ curves in Fig. 2(a) and (b) follow the Shockley *p-n* junction diode equation, after accounting for the voltage drop across the series resistance and the effects due to defect-mediated recombination and generation. We use the following equation to analyze the devices:

$$I_{DS} = I_0 \left( e^{q(V_{DS} - I_{DS} R_S)/\eta k_B T} - 1 \right) \tag{1}$$

where $I_0$ is the reverse saturation current, $R_S$ the series resistance, $\eta$ the diode ideality factor, $q$ the electron charge, $k_B$ the Boltzmann's constant, and $T$ the temperature. By fitting the forward bias current in Fig. 2(a) to Eq. 1, we extract $I_0$, $\eta$ and $R_S$; the best fits are shown as solid lines. The ideality factor provides a measure of the electron-hole recombination in the junction region; usually, $\eta = 1$ implies negligible recombination in the junction region, whereas $\eta = 2$ signifies defect-mediated recombination in the junction region. The relatively high values of $\eta$ we extract for the devices (1.6 for the 62 nm and 2.1 for the 18 nm flake) indicate a significant electron-hole recombination from defect levels. This is not unlike our previous observation where the same SG structure was used to create *p-n* junction diodes along individual single-walled carbon nanotubes [36]. There, we observed nearly ideal diode behavior ($\eta = 1$) only after suspending the nanotube in air over the junction region, suggesting that defect states are induced from the $SiO_2$ substrate. We expect to observe a similar trend with $MoS_2$, but suspending $MoS_2$ is beyond the scope of this work.

It is easy to verify that the rectifying $I_{DS}$-$V_{DS}$ characteristics shown in Fig. 2 are due to the formation of a *p-n* junction within $MoS_2$, and not due to the metal-$MoS_2$ Schottky barriers. We can separate the contribution of the *p-n* junction from those of the contacts by comparing the $I_{DS}$-$V_{DS}$ characteristics under asymmetric and symmetric doping configurations. With high symmetric SG biases, no potential barrier should form inside the $MoS_2$ channel, and the conductance should be governed by the two series connected metal-$MoS_2$ Schottky contacts. We confirm this in the insets of Fig. 2(b) by showing the characteristics under symmetric *n*- and *p*-doping. The $I_{DS}$-$V_{DS}$ curves there show only a small nonlinearity, which is more pronounced under *p-p* configurations. The substantially linear characteristics suggest that the metal-semiconductor contacts do not contribute to the highly rectifying characteristics of the device in the *p-n* doping configurations. As a further confirmation that the rectifying characteristics are due to the formation of a *p-n* junction, we note that the contact resistances $V_{DS}/I_{DS}$ from the insets of Fig. 2(b) are close to the extracted $R_S$ from the *p-n* diode *I-V* characteristics in Fig. 2(a), i.e. if the voltage drops due to the *p-p* and *n-n* resistances found in Fig. 2(b) insets are subtracted from the applied bias in Fig. 2(a), exponential *I-V* characteristics for the whole bias range are obtained, characteristic of a *p-n* junction diode.

The *p-n* diode we fabricate is fundamentally different from bulk diodes in one way: because the doping is achieved electrostatically, a depletion region does not form. But, since asymmetric carrier density is still present, a built-in voltage exists at the junction that can be used to separate photogenerated electron-hole pairs. Therefore, under illumination, the two *p-n* diodes in Fig. 2 show the photovoltaic effect, as shown in Fig. 3. The photovoltaic effect is characterized by a bias region where there is power gain, i.e. $I_{DS}V_{DS} < 0$. The important parameters characterizing the photovoltaic effect are the open-circuit voltage $V_{OC}$, the short-circuit current $I_{SC}$, the voltage $V_M$ and current $I_M$ at the maximum photogenerated power, and the fill factor (FF) defined as the ratio $V_M I_M / V_{OC} I_{SC}$ [37]. Compared to the 62 nm flake, the 18 nm thick device shows a lower $I_{SC}$ but a higher $V_{OC}$, for an overall improved photovoltaic effect. In applications of the photovoltaic effect, e.g. solar cells, a "square" like *I-V* profile is desirable which is quantified by the FF, with a value of 1 indicating a completely square profile with maximum photogenerated power in the device irrespective of bias. The FF for the 18 nm flake, which has a more "square"-like profile, is found to reasonably high at 0.63, while being significantly reduced for the 62 nm device (0.32). The reason for the low FF for the thicker device, despite having a much larger photocurrent, is likely due to the fact that, as pointed out earlier, in thicker flakes the screening of the gate-induced charges creates a region that shunts the *p-n* diode. Incidentally, this region is also the region that absorbs the most light and is expected to be less resistive under illumination. When modeled as a shunt resistor, the effect on the photovoltaic properties is to reduce the FF, as evident in Fig. 3 (bottom).

For devices that are only a few monolayers thick, while we found difficulties in demonstrating rectifying *p-n* diode *I-V* characteristics owing to the lack of ambipolar conduction (Fig. 1), it is nevertheless possible to observe the photovoltaic effect in these devices through creating a carrier density gradient. In the *p-n* configuration, for example, while the D-S current under bias might be too low to detect because of a high *p*-Schottky barrier, a short-circuit D-S current can still be generated under illumination, as the electric fields due to the carrier density variation sweep the photogenerated carriers across the barrier. Photogeneration also occurs at the Schottky junctions due to the built-in electric fields at the metal contacts. Under unipolar doping conditions, the photocurrents at the source and drain oppose each other. When a carrier density gradient, e.g. a *p-n* doping profile, is created, the built-in potentials at the Schottky junctions become asymmetric, and the photogeneration current in each might add to or oppose the one due to the carrier density gradient, depending on the doping profile. The sum of these three components is typically non-zero, leading to a D-S photocurrent. We observe such short-circuit photocurrents in all our *p-n* doped $MoS_2$ devices, the spectral response of which confirms the predicted band-gap change with thickness, as shown in Fig. 4. Here, the short-circuit current was measured by dispersing a broadband light source (quartz-tungsten-halogen lamp) through a monochromater (Horiba-JobinYvon iHR320) using a diffraction grating. The slit widths we used achieved monochromatic light with < 5 nm bandwidth. The measured short-circuit current $I_{SC}$ was normalized with respect to the photon flux, which was determined by calibrated photodiodes. The normalized $I_{SC}$ for monolayer, bilayer, and bulk $MoS_2$ devices are plotted as a function of incident photon energy in Fig. 4. They are offset vertically for clarity. For all the three devices, the onset of the direct band gap transition is quite evident with distinct peaks at ~1.9 and 2.1 eV, which correspond well to the previously observed absorption data in $MoS_2$, attributed to the "A" and "B" excitonic transitions [13], [14]. For the bulk device, there is a significant contribution to the photocurrent due to indirect gap transitions below 1.9 eV. We determine the direct gap transition energy for the three devices by fitting a Lorentzian peak function to each of the measured data at the "A" position; the fits are shown as solid lines in Fig. 4. The extracted values for this transition energy reduce from 1.97 eV for the monolayer device to 1.932 eV for bulk, a shift of $>4k_BT$ at 100 K. Overall, the energies of the "A" peak are higher than the values reported in literature [13, 14], and could arise from doping in our devices; the "A" excitonic energy in monolayer $MoS_2$ has been previously observed to increase with electron density and the consequent decrease in the exciton binding energy due to electrostatic shielding of the excitonic coulomb interactions [38]. The shift in the excitonic peak in our devices suggests an increase in the electrostatic shielding with decreasing layer thickness. We estimate the maximum electron/hole modulation by the SG biases in our devices to be $4 \times 10^{12}$ cm$^{-2}$ using a parallel-plate capacitance model, which, compared to the "A" peak shift trend in [38] can't account for the high value of 1.97 eV in our monolayer device. Therefore, the background electron doping in our devices must be considerably higher than the doping induced by the SG, likely due to induction from charged impurities from the substrate or adsorbates, which is consistent with the $n$-type doping we observe in the transfer characteristics at zero SG bias. For the bilayer and bulk device, we note that because of the increasing dimensionality, the induced charge can no longer be described by a sheet density, and the actual electron density, for the same SG bias, or charged impurity density, will be less than that for the monolayer device. Because of the reduced electron density, the electrostatic shielding of the exciton coulomb interaction will also be less, increasing the exciton binding energy, which could explain the red-shift in the exciton peak. Another reason for the shift could be the phonon-assisted indirect transition processes for the bilayer and bulk devices. Apart from the "A" and "B" excitonic peaks, additional features are observed at higher energies of the spectrum, especially for the mono- and bilayer devices, possibly due to the variation in the density of states in different branches of the valence and conduction bands.

Quantum efficiency as a function of energy for our diodes can be calculated using the relation $I_{SC}/qFA$, where $F$ is the photon flux and $A$ is area of the optically active region. We use the entire exposed area of the MoS$_2$ flake to calculate the efficiency. The calculated efficiency values of a few tenths of a percent, shown in Fig. 4, are lower than the observed photogeneration power conversion efficiency at Schottky junctions in bulk MoS$_2$, which are 1-2.5% [35], [39]. A reason for the low conversion efficiencies in our devices could be the fact that compared to these reports, in our devices there is an additional built-in field due to the carrier density gradient, the photocurrent due to which may oppose that due to the Schottky junctions. With careful choice of metal work functions and the levels of the *p* and *n* doping levels, it might be possible to optimize these photocurrent components to increase the total device current and the conversion efficiency. We note that the actual efficiency could be significantly higher as the area of the optically active region is likely to be considerably smaller than the total flake area that we use in the calculation.

In summary, reconfigurable electrostatic doping through buried gates is demonstrated in exfoliated MoS$_2$ flakes that allow studying the properties of *p-n* junction diodes and the photovoltaic effect. In MoS$_2$ flakes thicker than 10 nm, both electron and hole conduction are significant, enabling the creation of *p-n* diodes that show rectifying *I-V* characteristics and the photovoltaic effect. For a few monolayer-thick devices, while negligible hole conduction prevents measuring the *I-V* properties of a *p-n* junction, the photovoltaic effect can be still observed through the creation of a carrier density gradient across the device. The spectral response of the photocurrent shows the characteristic transition energies of the MoS$_2$ bandstructure, and a blue-shift for the direct gap transition with decreasing layer thickness.

**REFERENCES**

[1] K. Novoselov, A. Geim, S. Morozov, D. Jiang, Y. Zhang, S. Dubonos, I. Grigorieva, and A. Firsovet. *Science*, **306** (5696):666–669, 2004.

[2] K. Novoselov, A. Geim, S. Morozov, D. Jiang, M. Katsnelson, I. Grigorieva, S. Dubonos, and A. Firsov. *Nature*, **438** (7065):197–200, 2005.

[3] Y. Zhang, Y.-W. Tan, H. Stormer, and P. Kim. *Nature*, **438** (7065):201–204, 2005.

[4] K. Novoselov, D. Jiang, F. Schedin, T. Booth, V. Khotkevich, S. Morozov, and A. Geim. *Proc. Nat. Acad. Sci.,* **102** (30):10451–10453, 2005.

[5] S. Kim. A. Konar, W.-S. Hwang, J.-H. Lee, J. Lee, J. Yang, C. Jung, H. Kim, J.-B. Yoo, J.-Y. Choi, *et al. Nat Comm.*, **3**:1011, 2012.

[6] B. Radisavljevic, A. Radenovic, J. Brivio, V. Giacometti, and A. Kis. *Nat Nano.*, **6**:147–150, 2011.

[7] J. Pu, Y. Yomogida, K.-K. Liu, L.-J. Li, Y. Iwasa, and T. Takenobuet. *Nano lett.*, **12** (8):4013–4017, 2012.

[8] L. Britnell, R. Gorbachev, R. Jalil, B. Belle, F. Schedin, A. Mishchenko, T. Georgiou, M. Katsnelson, L. Eaves, S. V. Morozov, *et al*. *Science*, **335** (6071):947–950, 2012.

[9] W. J. Yu, Z. Li, Ha. Zhou, Y. Chen, Y. Wang, Y. Huang, and X. Duan. *Nat. mat.,* **12** (3):246–252, 2012.

[10] M. S. Choi, G.-H. Lee, Y.-J. Yu, D.-Y. Lee, S. H. Lee, P. Kim, J. Hone, and W. J. Yoo. *Nat. comm.,* **4**:1624, 2013.

[11] S. Bertolazzi, D. Krasnozhon, and A. Kis. *ACS nano*, **7** (4):3246–3252, 2013.

[12] D. Jariwala, V. Sangwan, C.-C. Wu, P. L. Prabhumirashi, M. Geier, T. Marks, L. Lauhon, and M. Hersam. *Proc. Nat. Acad. Sci.,* **110** (45):18076–18080, 2013.

[13] A. Splendiani, L. Sun, Y. Zhang, T. Li, J. Kim, C.-Y. Chim, G. Galli, and F. Wang. *Nano lett.,* **10** (4):1271–1275, 2010.

[14] K. F. Mak, C. Lee, J. Hone, J. Shan, and T. Heinz. *Phys. Rev. Lett.*, **105**:136805, Sep 2010.

[15] Z. Yin, H. Li, H. Li, L. Jiang, Y. Shi, Y. Sun, G. Lu, Q. Zhang, X. Chen, and H. Zhang. *ACS nano*, **6** (1):74–80, 2011.

[16] W. Choi, M. Y. Cho, A. Konar, J. H. Lee, G.-B. Cha, S. C. Hong, S. Kim, J. Kim, D. Jena, J. Joo, and S. Kim. *Adv. mat.*, **24** (43):5832–5836, 2012.

[17] Q. Wang, K. Kalantar-Zadeh, A. Kis, J. Coleman, and M. Strano. *Nat Nano.*, **7**: 699 – 712, 2012.

[18] D. Xiao, G.-B. Liu, W. Feng, X. Xu, and W. Yao. *Phys. Rev. Lett.*, **108**:196802, May 2012.

[19] H. Zeng, J. Dai, W. Yao, D. Xiao, and X. Cui. *Nat nano*, **7**(8), 490-493, 2012.

[20] K. Mak, K. He, J. Shan, and T. Heinz. *Nat nano*, **7**(8), 494-498, 2012.

[21] T. Cao, G. Wang, W. Han, H. Ye, C. Zhu, J. Shi, Q. Niu, P. Tan, E. Wang, B. Liu, and J. Feng. *Nat comm.,* **3**: 887, 2013.

[22] M. Chen, H. Nam, S. Wi, L. Ji, X. Ren, L. Bian, S. Lu, and X. Liang. *Appl. Phys. Lett.*, **103** (14):142110, 2013.

[23] Y. Zhang, J. Ye, Y. Matsuhashi, and Y. Iwasa. *Nano lett.,* **12** (3):1136–1140, 2012.

[24] Y. Zhang, J. Ye, Y. Yomogida, T. Takenobu, and Y. Iwasa. *Nano lett.*, **13** (7):3023–3028, 2013.

[25] S. Sutar, J. Liu, E. Comfort, T. Taniguchi, K. Watanabe, and J. U. Lee. *Nano Lett.*, **12** (9):4460–4464, 2012.

[26] C. Lee, H. Yan, L. E. Brus, T. F. Heinz, J. Hone, and S. Ryu. *ACS Nano*, **4**(5):2695–2700, 2010.

[27] W. Bao, X. Cai, D. Kim, K. Sridhara, and M. Fuhrer. *Appl. Phys. Lett.*, **102**:042104, 2013.

[28] S. Tans, A. Verschueren, and C. Dekker. *Nature*, **393** (6680):49–52, 1998.

[29] R. Martel, T. Schmidt, H. Shea, T. Hertel, and P. Avouris. *Appl. Phys. Lett.,* **73:2447**, 1998.

[30] V. Derycke, R. Martel, J. Appenzeller, and P. Avouris. *Appl. Phys. Lett.*, **80**:2773, 2002.

[31] V. Derycke, R. Martel, J. Appenzeller, and P. Avouris. *Nano Lett.*, **1**(9):453–456, 2001.

[32] R. Martel, V. Derycke, C. Lavoie, J. Appenzeller, K. K. Chan, J. Tersoff, and P. Avouris. *Phys. Rev. Lett.*, **87** (25):256805, 2001.

[33] M. Bockrath, J. Hone, A. Zettl, P. L. McEuen, A. G. Rinzler, and R. E. Smalley. *Phys. Rev. B*, **61**:R10606–R10608, 2000.

[34] C. Dean, A. Young, I. Meric, C. Lee, L. Wang, S. Sorgenfrei, K. Watanabe, T. Taniguchi, P. Kim, K. Shepard, J. Hone. *Nat Nano*, **5**(10):722–726, 2010.

[35] M. Fontana, T. Deppe, A. Boyd, M. Rinzan, A. Liu, M. Paranjape, and P. Barbara. *Scientific reports* **3**: 1634 (2013).

[36] J. U. Lee, P. Gipp, and C. Heller. *Appl. Phys. Lett.*, **85** (1):145–147, 2004.

[37] M. Green. *Solar cells: operating principles, technology, and system applications*, vol. 1. Englewood Cliffs, NJ, Prentice-Hall, Inc. 1982.

[38] K. F. Mak, K. He, C. Lee, G. H. Lee, J. Hone, T. Heinz, and J. Shan. *Nat. Mat.*, **12**(3):207–211, 2012.

[39] E. Fortin, and W. Sears. *J. Phys. Chem. Sol.,* **43**(9): 881-884, 1982.

**FIGURES**

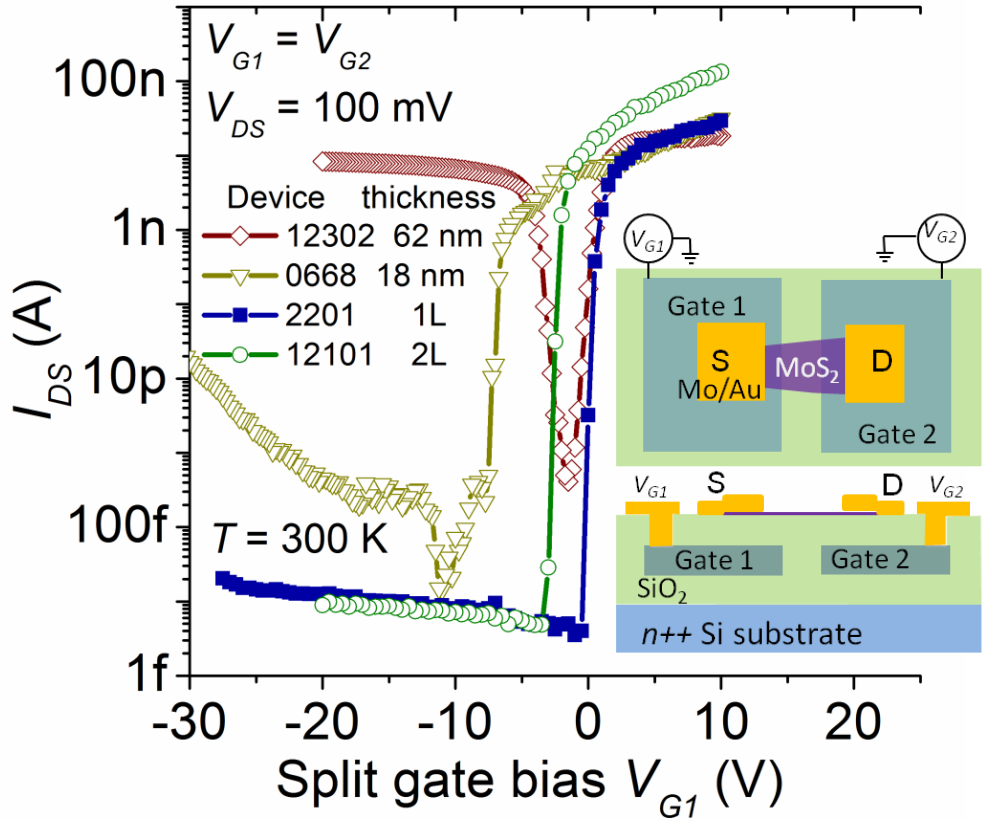

**Figure 1.** Two-terminal electrical transfer characteristics of exfoliated $MoS_2$ flakes: inset shows schematics of the device structure. Main plot shows the D-S current as a function of the SG biases (equal to each other). For thicker flakes, strong ambipolar conduction is observed whereas the mono- and bilayer devices show only *n*-type conduction.

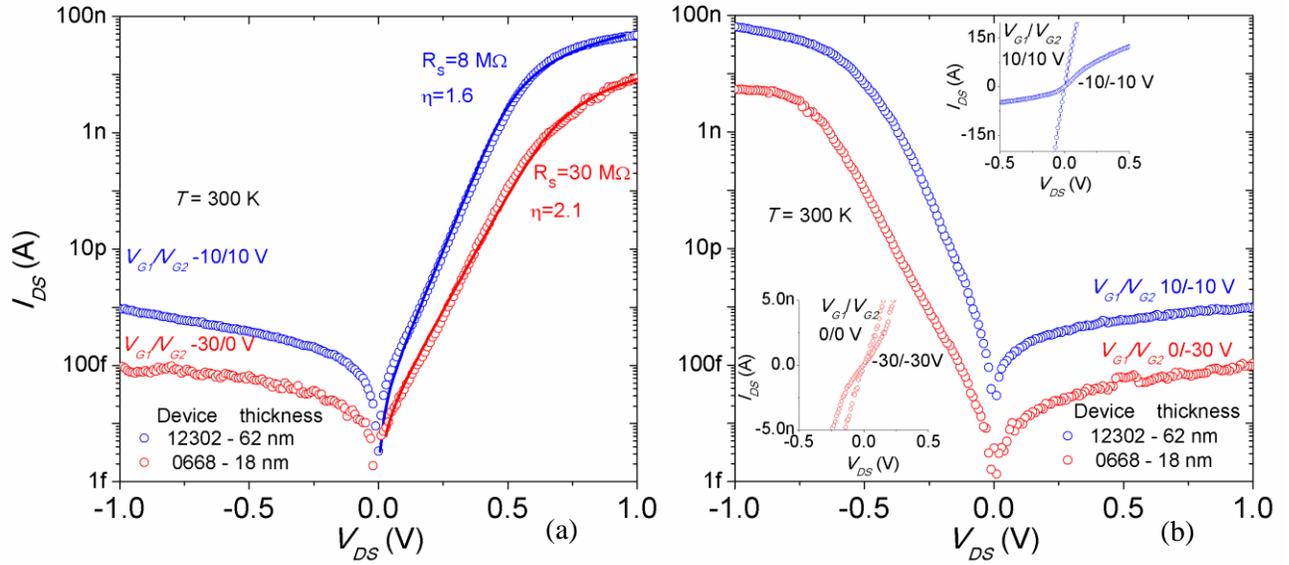

**Figure 2.** Source-drain $I_{DS}$-$V_{DS}$ measurements of junctions formed in MoS$_2$ flakes, we plot the magnitudes of $I_{DS}$: (a) the SGs are biased to create a *p-n* junction between the drain and source; markers show the measured characteristics; lines are fits to the data using the diode equation; (b) the bias polarities on the SGs are changed while keeping the same magnitudes to create an *n-p* (main plot), and *p-p, n-n* junctions (insets) between the drain and source. The main plot of (b) is almost a mirror image of that in (a), showing reconfigurability of the electrostatic doping. For the *p-p* and *n-n* doping, the *I-V* characteristics show linear behavior and no rectification.

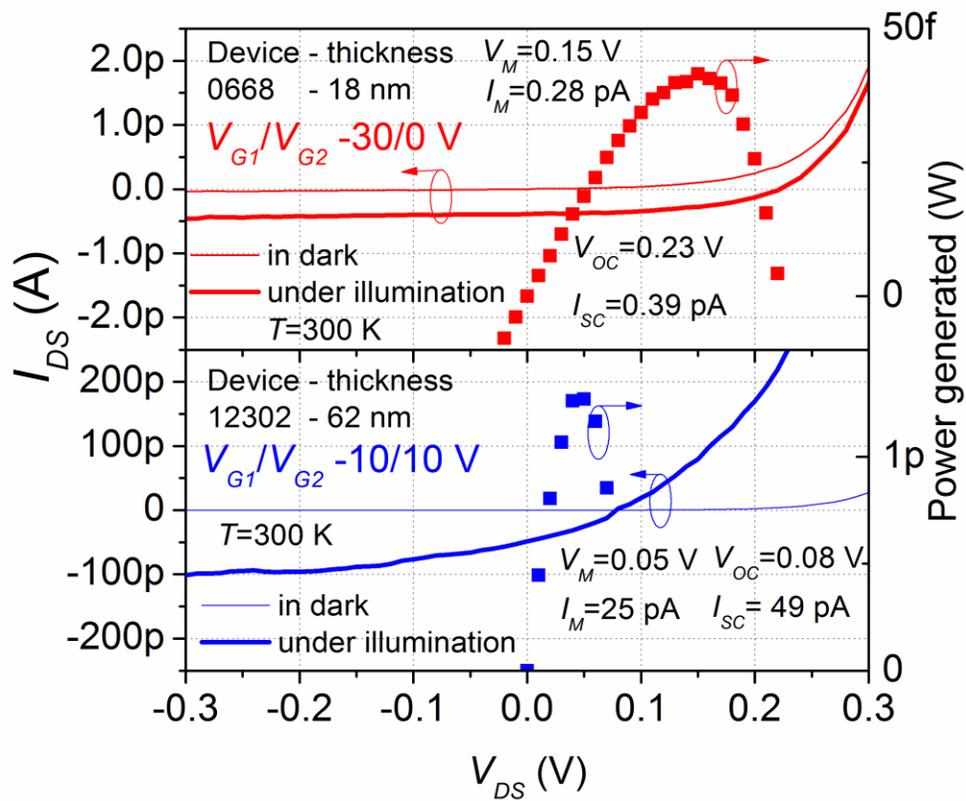

**Figure 3.** Photovoltaic effect in MoS$_2$ *p-n* diodes: *I-V* characteristics under illumination show an open-circuit voltage and a short-circuit current; for a range of biases the product $I_{DS}V_{DS}$ becomes negative, indicating photovoltaic power generation.

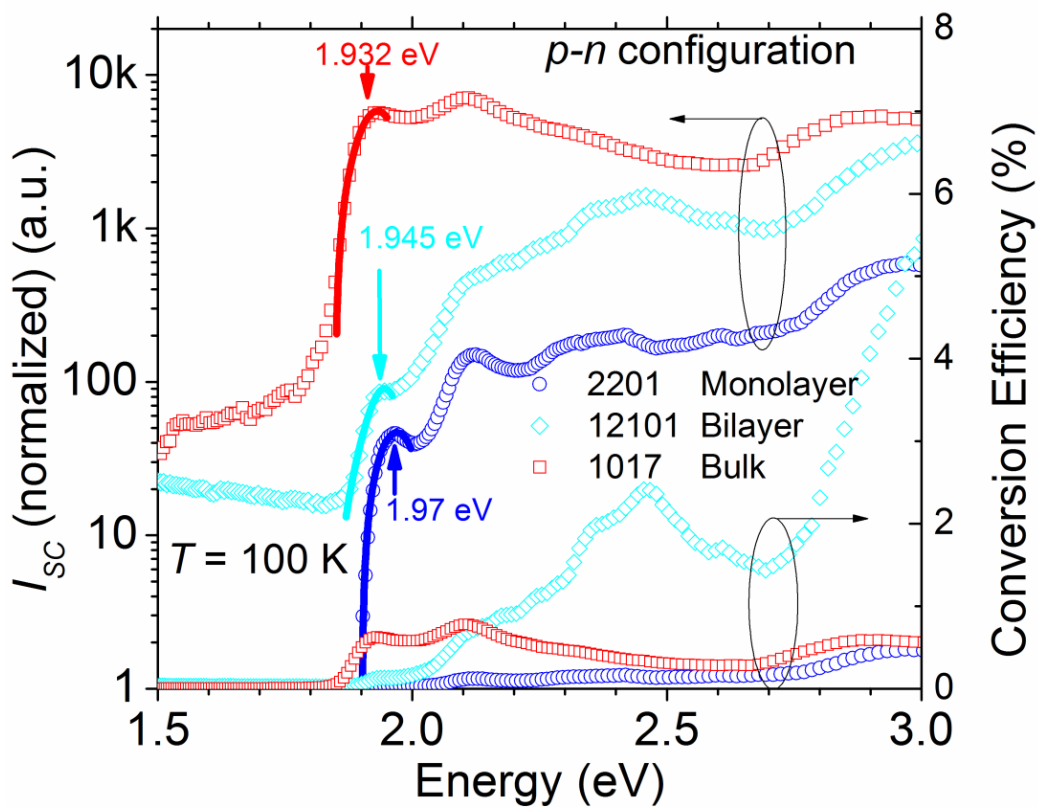

**Figure 4.** Short-circuit photocurrent spectrum and photon conversion efficiency of monolayer, bilayer and bulk $MoS_2$ devices. The photocurrent shows characteristic transitions with the energy band gap. The currents have been vertically offset for clarity. The photon conversion efficiency of the devices are found from the absolute values of the photocurrent, assuming that light absorption in the entire flake contributes to the photocurrent.